\documentclass[epj]{svjour}
\usepackage{textcomp}
\usepackage{graphics}
\usepackage{url}
\begin{document}
\title{Cosmic-ray muon flux at Canfranc Underground Laboratory}
\subtitle{}

\author{
Wladyslaw~Henryk~Trzaska\inst{1} \and 
Maciej~Slupecki\inst{1} \and
Iulian~Bandac\inst{2} \and 
Alberto~Bayo\inst{2} \and 
Alessandro~Bettini\inst{2} \and 
Leonid~Bezrukov\inst{3} \and 
Timo~Enqvist\inst{1}\textsuperscript{,}\inst{4} \and 
Almaz~Fazliakhmetov\inst{3}\textsuperscript{,}\inst{5} \and 
Aldo~Ianni\inst{2} \and Lev Inzhechik\inst{5} \and 
Jari~Joutsenvaara\inst{4} \and 
Pasi~Kuusiniemi\inst{1}\textsuperscript{,}\inst{4} \and 
Kai~Loo\inst{1} \and 
Bayarto~Lubsandorzhiev\inst{3} \and 
Alexander~Nozik\inst{3}\textsuperscript{,}\inst{5} \and 
Carlos~Pe\~na~Garay\inst{2} \and 
Maria~Poliakova\inst{3}\textsuperscript{,}\inst{5}
}                     
%
%
\institute{
Department of Physics, University of Jyv\"askyl\"a, Finland \and 
Laboratorio Subterraneo de Canfranc, Spain \and 
Institute of Nuclear Research, Russian Academy of Sciences, Moscow, Russia \and
Kerttu Saalasti Institute, University of Oulu, Finland \and
Moscow Institute of Physics and Technology, Russia
}
\date{Received: date / Revised version: date}
%
\abstract{
Residual flux and angular distribution of high-energy cosmic muons have been measured in two underground locations at the Canfranc Underground Laboratory (LSC) using a dedicated Muon Monitor. The instrument consists of three layers of fast scintillation detector modules operating as 352 independent pixels. The monitor has a flux-defining area of 1~m${}^{2}$ and covers all azimuth angles, and zenith angles up to 80\textdegree. The measured integrated muon flux is \mbox{$(5.26 \pm 0.21) \times 10^{-3}$~m${}^{-2}$s${}^{-1}$} in the Hall A of the LAB2400 and \mbox{$(4.29 \pm 0.17) \times 10^{-3}$~m${}^{-2}$s${}^{-1}$} in LAB2500. The angular dependence is consistent with the known profile and rock density of the surrounding mountains. In particular, there is a clear maximum in the flux coming from the direction of the Rioseta valley.
\PACS{
      {29.40.Mc}{Scintillation detectors}   \and
      {95.85.Ry}{Neutrino, muon, pion, and other elementary particles; cosmic rays}
     } 
} 
\maketitle

\section{Introduction}
\label{sec:1}
Reduction of the intense particle flux induced by cosmic rays is one of the main reasons to locate low-background laboratories underground. Consequently, the residual muon intensity is a key parameter in site selection and evaluation. While the processes of creation of particle showers and their transport through the atmosphere and through the layers of rock and sediments are relatively well understood, the detailed numerical data on the geological structures above and around the laboratory are seldom available with the desired accuracy. In the end, a direct measurement is the best way to determine precisely the actual muon flux at the given underground location. For a comprehensive review of the deep underground laboratories and their scientific projects see the contributions to the focus issue of the European Physical Journal Plus 127~(2012)~\cite{1_Bertou:2012fk,2_Chen:2012fm,3_Mondal:2012fn,4_Lesko:2012fp,5_Smith:2012fq,6_Votano:2012fr,7_Piquemal:2012fs,8_Suzuki:2012ft,9_Bettini:2012fu,10_Kuzminov:2012fv,11_Bettini:2012fw}. Several dedicated measurements of cosmic muons in various underground laboratories are described in~\cite{12_Enqvist:2005cy,13_Bandac:2017jjm,14_Gray:2010nc,15_Yu-Cheng:2013iaa,16_Robinson:2003zj,17_Blyth:2015nha,18_Kalousis:2014wba,19_Esch:2004zj,Schmidt:2013gdc}.

\begin{figure*}
\centering
\resizebox{1.0\textwidth}{!}{
  \includegraphics{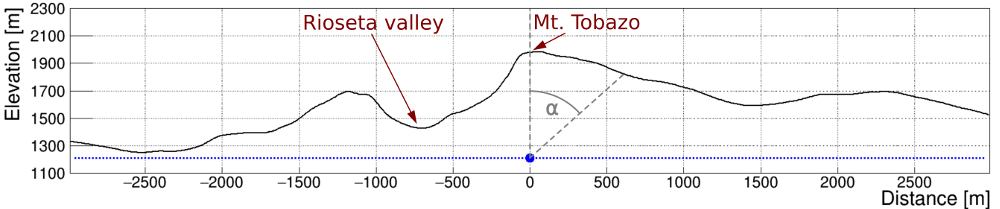}
}
\caption{Mountain profile along the path of the railroad tunnel. The distances are given from the Hall A of LAB2400, marked as a large dot at $x = 0$~m. The positive direction is towards France. The blue dotted line indicates the elevation of the floor level at the LSC.}
\label{fig:1:mntprofile}
\end{figure*}

The Canfranc Underground Laboratory (LSC)~\cite{20_Ianni:2016fjt} is located under the Mount Tobazo (1980~m) in the Aragonese Pyrenees. The laboratory caverns have been excavated between the vacant train tunnel and the modern road tunnel joining Spain and France. Both tunnels are used as access routes to the laboratory area. The coordinates of the LSC are known with accuracy of $\pm5\mathrm{~cm}$. 
The exact position of the Muon Monitor in the LAB2400 was: floor altitude 1204.48 m above the sea
level, the longitude 0\textdegree{}~31'~44.85570"~W, and the latitude 42\textdegree{}~46'~28.99971"~N. In the LAB2500 the corresponding values were: 1206.47~m, 0\textdegree{}~31'~45.26066"~W, 42\textdegree{}~46'~31.04089"~N.

\begin{figure}
\centering
\resizebox{0.5\textwidth}{!}{
  \includegraphics{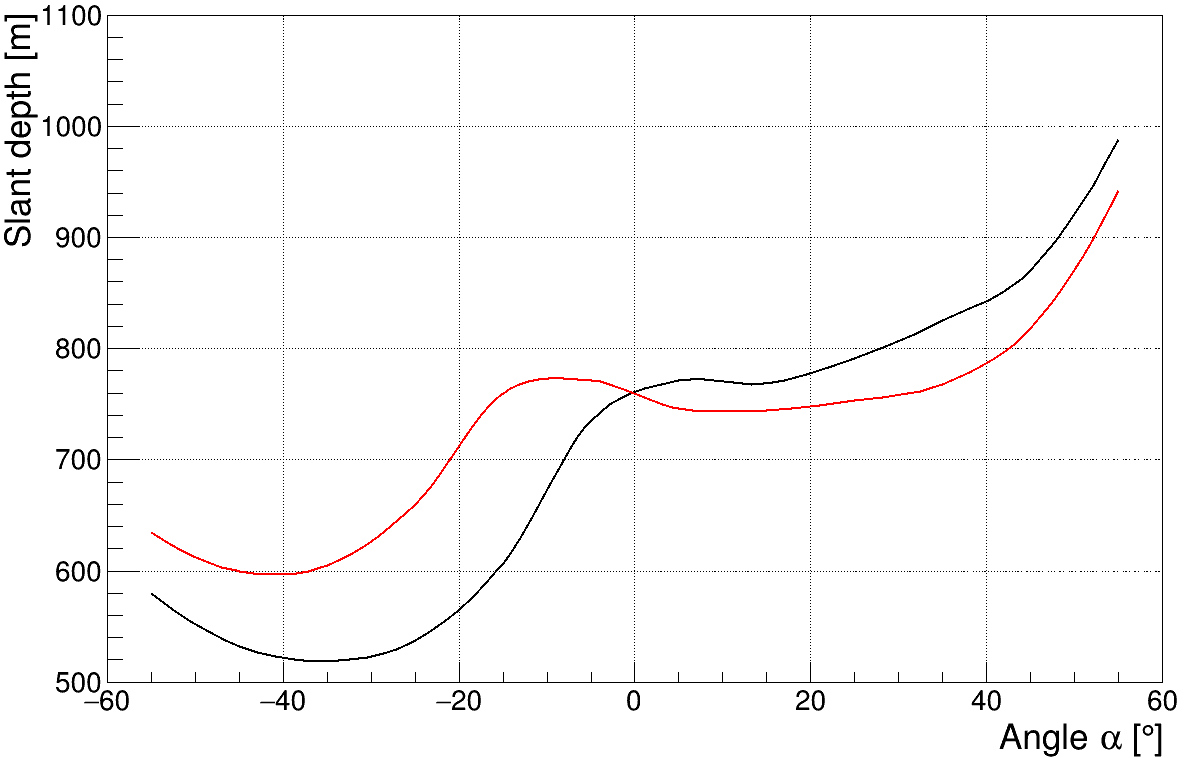}
}
\caption{Slant depth as a function of the projection angle along the plane defined by the railroad tunnel. The black line corresponds to the view from LAB2400. The red line, from LAB2500. The positive angles are in the SSE direction (towards France). The negative, in the NNW direction (towards Spain).}
\label{fig:2:profiledeg}
\end{figure}

Fig.~\ref{fig:1:mntprofile} shows the cross section of the mountain range along the railway tunnel, following the SSE$-$NNW direction. Mount Tobazo, situated directly over the Hall A of LAB2400 (distance = 0 m), is the highest point. The valley of the Rioseta river, at minus 750 meters from Hall A, has the lowest elevation. These elevation changes in the profile of the mountain range surrounding LSC result in significant variations of the slant depth for different projection angle, as shown in Fig. 2. The projection angle alpha, defined in Fig. 1., is analogue to the zenith angle with the azimuth plane fixed to the SSE-NNW direction. The data for the plots in Fig.~\ref{fig:1:mntprofile} and Fig.~\ref{fig:2:profiledeg} were extracted from the April 2018 release of the dataset from the Advanced Land Observing Satellite (ALOS)~\cite{alos}.

\section{Experimental setup}
\label{sec:2}

\begin{figure}
\centering
\resizebox{0.43\textwidth}{!}{
  \includegraphics{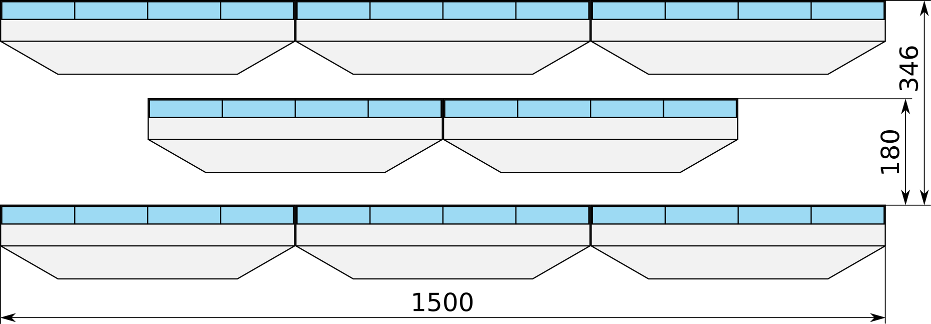}
}
\caption{Schematic view of the Muon Monitor. It consisted of 22 SC16 units. Each SC16 has 16 individual scintillator pixels, rendered in blue. The case of SC16 is outlined in grey. The dimensions are in mm.}
\label{fig:3:mm}
\end{figure}

The measurements were performed with a Muon Monitor (MM) assembled especially for that purpose.  It consisted of an array of 352 individual scintillator pixels arranged in 3 layers, as shown in Fig.~\ref{fig:3:mm}. The active volume of one scintillator pixel was \mbox{$122 \times 122 \times 30$~mm${}^3$}. The key building block of the MM setup was a SC16 module housing 16 individual scintillators/pixels in one sturdy steel box, 120~mm thick, \mbox{$500 \times 500$~mm${}^{2}$} at the base. The top and bottom layers of the MM were made of 9 SC16 elements each. The flux-defining middle layer was made of 4 SC16 units and had the active area of 0.95~m${}^{2}$. The maximum detectable zenith angle for this configuration was approximately 80\textdegree.

The SC16 detectors were originally designed and constructed for the EMMA experiment~\cite{21_Kuusiniemi:2018vbz} in the Pyh\"asalmi mine in Finland. The time resolution is around 1.5~ns~\cite{21_Kuusiniemi:2018vbz}. The intrinsic detection efficiency of SC16 for muons is 100\%. It means that only muons traversing less than the nominal 3 cm thickness of the active layer may avoid detection. The acceptance of the MM is thus defined by the geometry shown in Fig.~\ref{fig:3:mm}. The measured total efficiency of a single layer of SC16s, accounting for the gaps between the scintillator pixels, is 98\%. For a three-layer coincidence event, the efficiency is 94\%. The energy threshold was set at around 2~MeV. For comparison, a muon traversing the 3~cm thickness of a single scintillator layer, generates a signal of around 6~MeV. To generate a valid trigger, at least one pixel in each layer had to register an event with energy above 2~MeV. A typical trigger rate was around 20~triple-layer coincidences per hour. 

\begin{figure}
\centering
\resizebox{0.38\textwidth}{!}{
  \includegraphics{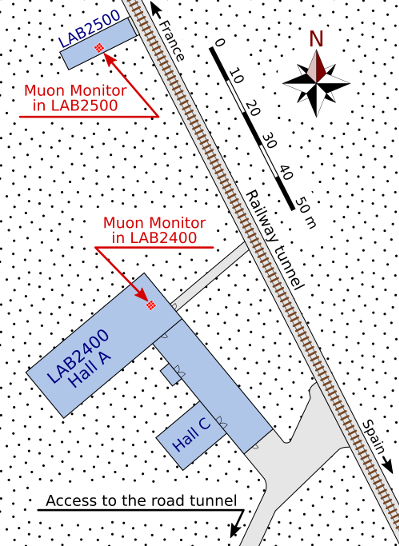}
}
\caption{Position of the Muon Monitor during the measurements at both locations.}
\label{fig:4:layout}
\end{figure}

Throughout the acquisition period all detector pixels remained active and the electronics operated in a stable way. As a result, the overall data quality is very good. In the Hall A the data were recorded from September 2013 till October 2015. The effective acquisition time was 584 days. In LAB2500 the measurements took place from October 2015 till March 2018 with the effective acquisition time of 569 days. The exact position of the MM at both locations is shown in Fig.~\ref{fig:4:layout}. 

\begin{figure}
\centering
\resizebox{0.5\textwidth}{!}{
  \includegraphics{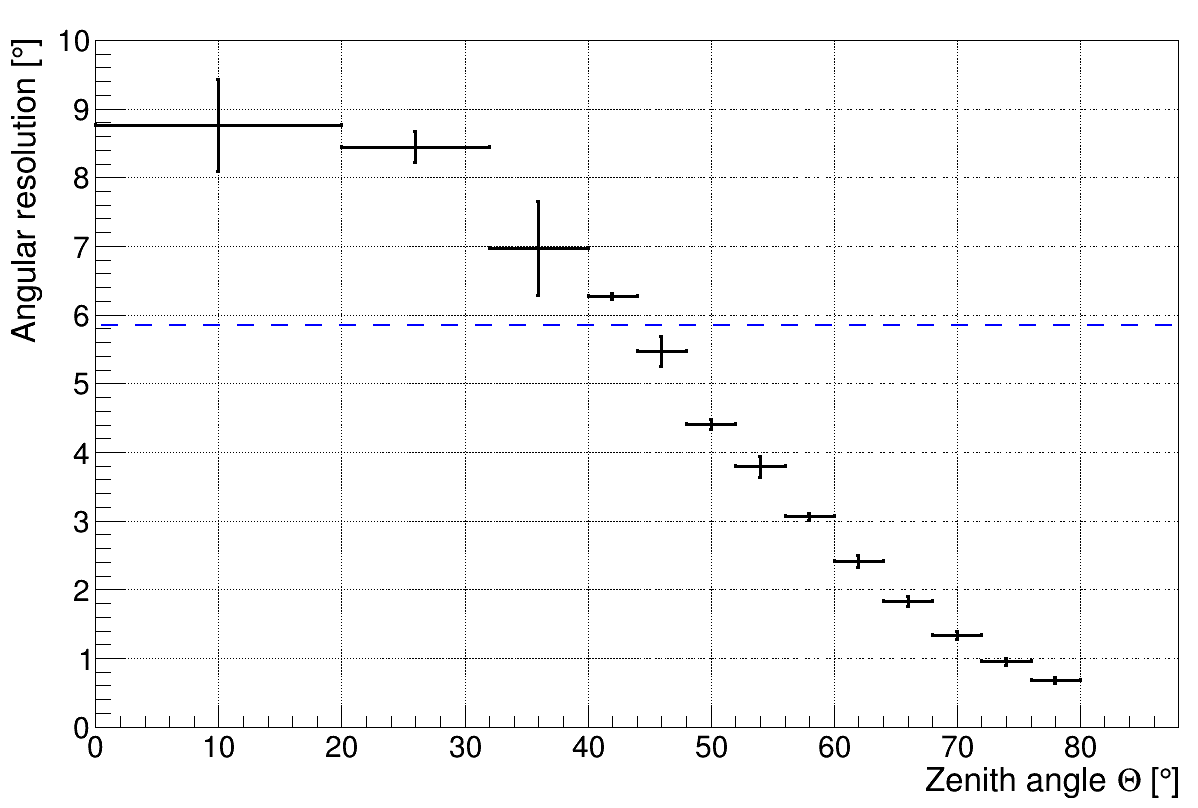}
}
\caption{Zenith angle resolution of the Muon Monitor as the function of the zenith angle. The blue dashed line represents the global resolution.}
\label{fig:5:thetares}
\end{figure}

\begin{figure}
\centering
\resizebox{0.5\textwidth}{!}{
  \includegraphics{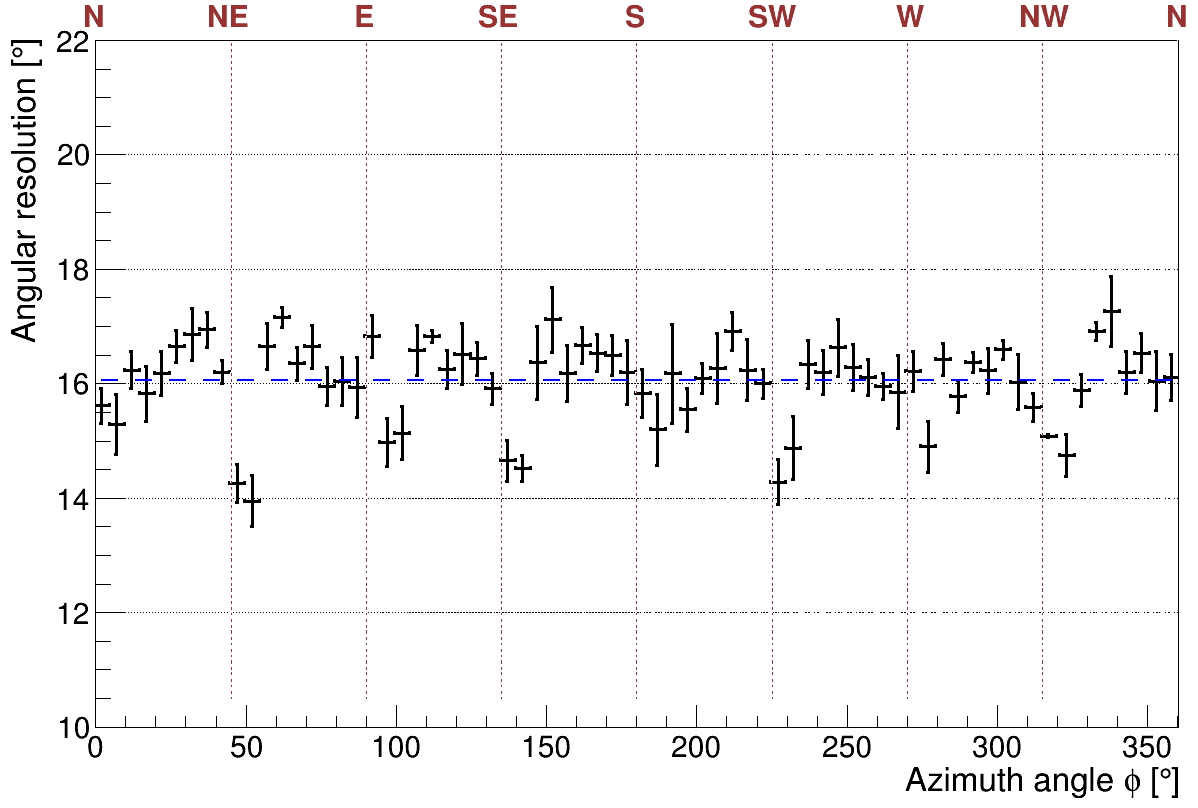}
}
\caption{Azimuth angle resolution of the Muon Monitor as the function of the azimuth angle. The blue dashed line represents the global resolution.}
\label{fig:6:phires}
\end{figure}

The angular resolution of the MM is a function of both the zenith angle (shown in Fig.~\ref{fig:5:thetares}) and the azimuth angle (shown in Fig.~\ref{fig:6:phires}). The dependence from $\phi$ comes from the square shape of the scintillator pixels yielding slightly better resolution when the azimuth angle is aligned with the sides of the scintillator squares and not with the diagonal directions. The $\Theta$ dependence comes from the fact that at larger zenith angles the projected pixel footprint gets smaller. For the numerical assessment of the angular resolution we have used a Monte Carlo approach. The firing pattern of the pixels from a simulated muon was reconstructed and the angular difference between the simulated and the extracted muon directions were compared. As the distributions were not a perfect Gauss curves, we have used the root mean square (RMS) instead of sigma as a parameter to characterise the angular resolution of the MM as a function of $\Theta$ and $\phi$.

\begin{figure*}
\centering
\resizebox{1.0\textwidth}{!}{
  \includegraphics{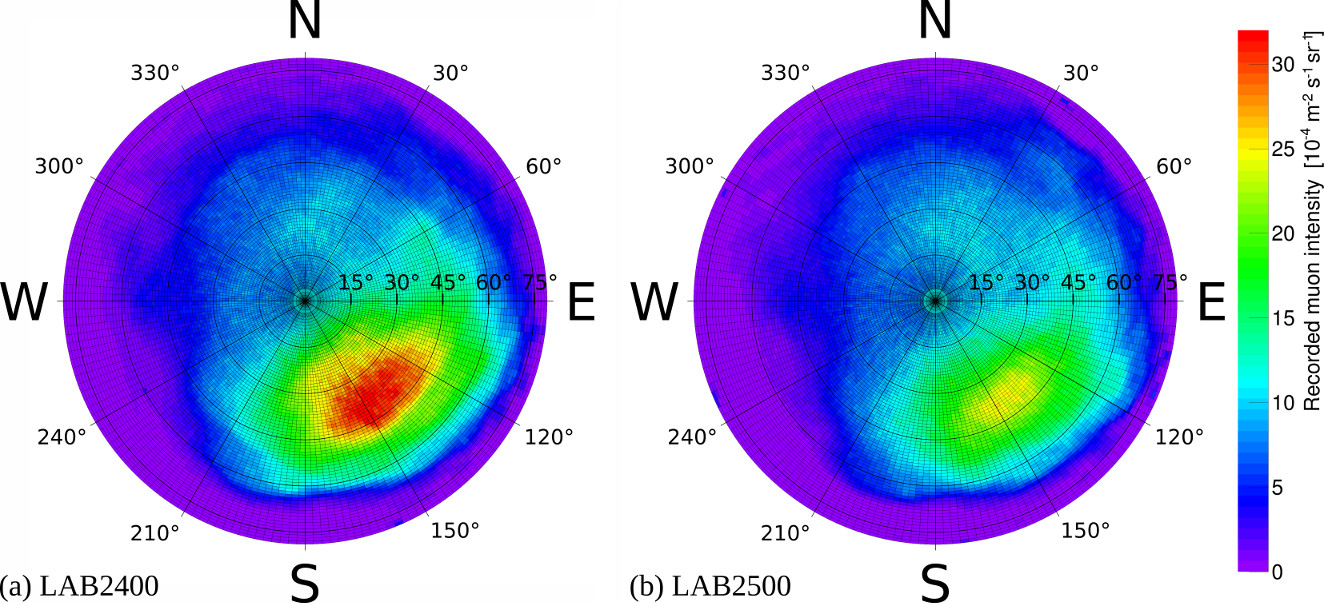}
}
\caption{Muon intensity measured in LAB2400 (left) and in LAB2500 (right) plotted as a function of the azimuth and zenith angle. The maxima measured around $\theta = 40$\textdegree~ and $\phi = 150$\textdegree~ point towards the Rioseta valley.}
\label{fig:7:angdistr}
\end{figure*}

Additional information about the experimental setup is provided in~\cite{13_Bandac:2017jjm}. For the description of the electronics, see~\cite{22_Yanin:2011fh,Volchenko}. The full details concerning the experimental setup, detectors, electronics, data acquisition, and data analysis will be described in a dedicated instrumental paper.

\section{Results and discussion}
\label{sec:3}

The integrated muon flux measured in the Hall A of the LAB2400 is \mbox{$(5.26 \pm 0.21) \times 10^{-3}$~m${}^{-2}$s${}^{-1}$}. The corresponding value for the LAB2500 is $(4.29 \pm 0.17) \times 10^{-3}$ m${}^{-2}$s${}^{-1}$. Because of the long duration of the measurements, needed to extract the angular distributions, the statistical fluctuations of the integrated fluxes are negligible ($\sim0.2\%$). The dominant uncertainty, estimated at $\pm 4$\%, is due to the systematics. 

Roughly 72\% of the registered events are single muons passing on a straight trajectory through the three layers of MM. However, the remaining 28\% of the events have a more complex pixel pattern. The uncertainty about the interpretation of these events is the main source of the systematic error of the extracted integrated muon flux. The probability of multiple muons passing simultaneously through the active area of MM is negligible. Nevertheless, muons generate electromagnetic (EM) showers while traversing, for instance, the celling of the cavern. Our preliminary GEANT4~\cite{geant4} simulations indicate that about $\frac{3}{4}$ of the \mbox{complex-pattern} events detected by MM contain both a muon hit and the EM component. The remaining $\frac{1}{4}$ are hits only by particles from EM showers and hence should be rejected from the integration of the muon flux. Further sources of error are dispersion in pixel efficiencies and alignment accuracy.

The angular distributions of the muon flux displayed as a function of the azimuth and zenith angle are shown in Fig.~\ref{fig:7:angdistr}. To produce the plot, only the single-muon data were used with unambiguously defined arrival angle. This subset represents about 72\% of the collected events. A two-step approach was needed to obtain continuous distributions from the coarsely-sampled data extracted from 352 pixels. First, the angular phase-space was determined for each pixel combination. Next, the registered coincidence was randomly allocated to one of the directions from the accessible phase-space for the given pixel sequence. The result is a smooth distribution with no detectable artefacts or remanences of the original pixelization.  As expected, for both locations the maximum flux is observed from the direction of the Rioseta valley, around the zenith angle of $\theta = 40$\textdegree~ and the azimuth angle of $\phi = 150$\textdegree.

\subsection{Consistency check}
\label{sec:3.1}

\begin{figure}
\centering
\resizebox{0.5\textwidth}{!}{
  \includegraphics{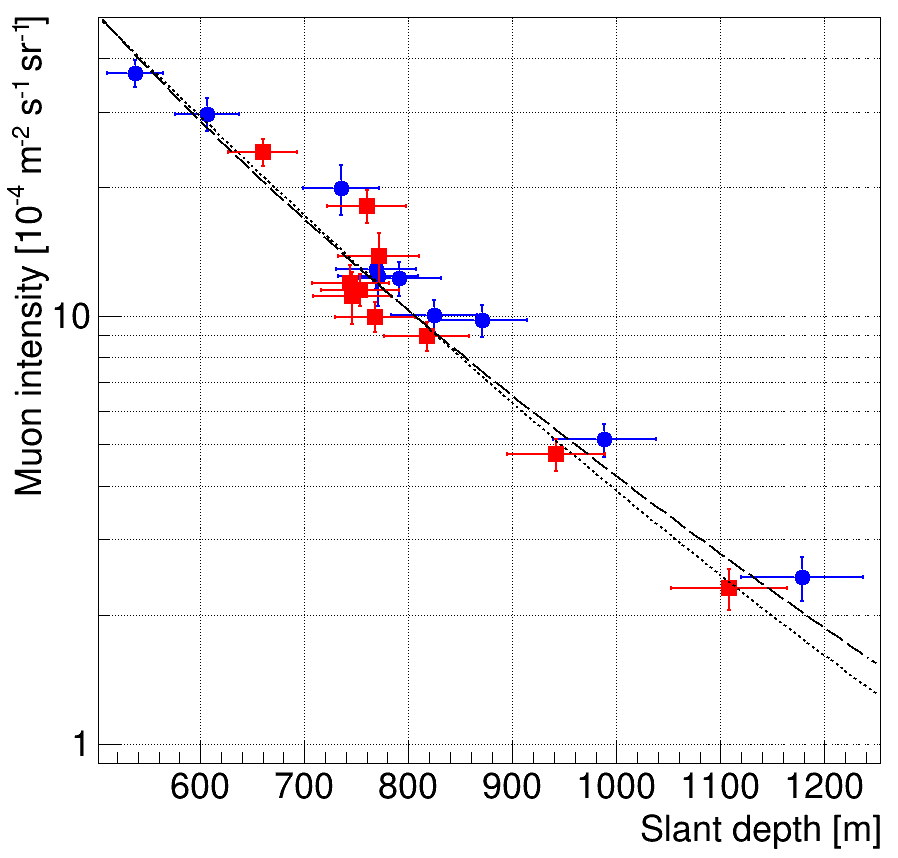}
}
\caption{
Muon intensity as a function of slant depth. The blue circles were measured at LAB2400 and the red squares, at LAB2500. The error bars represent 5\% uncertainty in the projected thickness determination from the satellite data and the 4\% systematic error of the measured intensity. The best fit using~(\ref{eq:1}) yielded  the average rock density of 2.67 g/cm${}^3$ (dashed curve), while~(\ref{eq:2}) yielded 2.73 g/cm${}^3$ (dotted curve).
}
\label{fig:8:corr}
\end{figure}

By combining the efficiency-corrected angular distribution of the muon flux from Fig.~\ref{fig:7:angdistr} with the satellite data on the shape of the terrain shown in Fig.~\ref{fig:2:profiledeg}, one can correlate the flux arriving from the given direction with the slant depth along that path. The cut along the railroad was selected for historical reasons and because it exhibits the most extreme variation in the slant depth.
The outcome is plotted in Fig.~\ref{fig:8:corr}. To obtain a similar result without muon tracking would require a series of measurements at multiple locations of different depth.

It has been pointed out~\cite{24_Lipari,25_Formaggio:2004ge} that, for the overburden values comparable to LSC, there is a simple relation between depth and muon intensity:

\begin{equation}
\label{eq:1}
I(x) \approx A \left( \frac{X_0}{x} \right) ^{\eta} e^{-\frac{x}{X_0}}
\end{equation}

Where, according to~\cite{25_Formaggio:2004ge}, fits to the existing data show $A = (2.15 \pm 0.08) \times 10^{-6}$~cm${}^{-2}$s${}^{-1}$sr${}^{-1}$, $\eta = 1.93^{+0.20}_{-0.12}$ and $X_0 = 1155^{+60}_{-30}$~m.w.e..

The authors of~\cite{Mei:2005gm} propose a different semi-empirical relation:

\begin{equation}
\label{eq:2}
I(x) \approx \left( I_{1}e^{-x/\lambda_{1}} + I_{2}e^{-x/\lambda_{2}} \right)
\end{equation}

Where, fits to the existing data show 
$I_{1} = (8.60 \pm 0.53) \times 10^{-6}$~cm${}^{-2}$s${}^{-1}$sr${}^{-1}$, 
$I_{2} = (0.44 \pm 0.06) \times 10^{-6}$ cm${}^{-2}$s${}^{-1}$sr${}^{-1}$,
$\lambda_{1} = (450 \pm 10)$~m.w.e. and
$\lambda_{2} = (870 \pm 20)$~m.w.e..

As a consistency check we have fitted both formulae to the data points from Fig.~\ref{fig:8:corr}. The slant depth, expressed in meters of water equivalent (m.w.e.) in~(\ref{eq:1}) and~(\ref{eq:2}), was converted into meters of rock by dividing it by a free parameter representing rock density. The best fit with~(\ref{eq:1}) yielded the average density of 2.67 g/cm${}^3$ (dashed curve) and 2.73 g/cm${}^3$ (dotted curve) with~(\ref{eq:2}). Both values are within 1\% from the expected density of limestone (2.7 g/cm${}^3$) that is the dominant component of the rock in the vicinity of LSC. This agreement confirms the consistency between the measured muon flux and the known geology and shape of the mountain above the LSC. The main reason for the relatively large horizontal error bars in Fig.~\ref{fig:8:corr} is the limited angular resolution of the MM.

\section{Summary and conclusions}
\label{sec:4}

The residual flux and angular distribution of high-energy cosmic muons in the Canfranc Underground Laboratory (LSC) have been measured. The integrated muon flux is $(5.26 \pm 0.21) \times 10^{-3}$~m${}^{-2}$s${}^{-1}$ for LAB2400 (Hall A) and $(4.29 \pm 0.17) \times 10^{-3}$~m${}^{-2}$s${}^{-1}$ for LAB2500. These results supersede the preliminary values published earlier~\cite{13_Bandac:2017jjm} where the shower-contaminated events were not included in the analysis and hence the old value was
underestimated by 20\% compared to the new result for LAB2400. For each site the data were collected over the period of nearly 600 days. The measurements were done with the Muon Monitor assembled especially for that purpose. The obtained angular dependence is consistent with the known mountain profile and rock density. In particular, there is a clear maximum in the flux from the direction of the Rioseta valley. As a result, the integrated muon flux is larger than what one would expect from the thickness of the overburden directly above the site. Consequently, some of the older evaluations have underestimated the integrated muon flux at LSC by up to a factor of two.

\section{Acknowledgments}
\label{sec:5}

We gratefully acknowledge the help and support from the directorate and staff of the Canfranc Underground Laboratory during the design, planning, construction and data acquisition phases of this project. This work has been supported in part by the Council of Oulu Region, the European Union Regional Development Fund, and by the grant of the Ministry of Education and Science of the Russian Federation number 3.3008.2017.

 \bibliographystyle{h-physrev}
 \bibliography{mm}

\begin{thebibliography}{10}

\bibitem{1_Bertou:2012fk}
X.~Bertou,
\newblock Eur. Phys. J. Plus {\bf 127}, 104 (2012).

\bibitem{2_Chen:2012fm}
H.-S. Chen,
\newblock Eur. Phys. J. Plus {\bf 127}, 105 (2012).

\bibitem{3_Mondal:2012fn}
N.~K. Mondal,
\newblock Eur. Phys. J. Plus {\bf 127}, 106 (2012).

\bibitem{4_Lesko:2012fp}
K.~T. Lesko,
\newblock Eur. Phys. J. Plus {\bf 127}, 107 (2012).

\bibitem{5_Smith:2012fq}
N.~J.~T. Smith,
\newblock Eur. Phys. J. Plus {\bf 127}, 108 (2012).

\bibitem{6_Votano:2012fr}
L.~Votano,
\newblock Eur. Phys. J. Plus {\bf 127}, 109 (2012).

\bibitem{7_Piquemal:2012fs}
F.~Piquemal,
\newblock Eur. Phys. J. Plus {\bf 127}, 110 (2012).

\bibitem{8_Suzuki:2012ft}
Y.~Suzuki and K.~Inoue,
\newblock Eur. Phys. J. Plus {\bf 127}, 111 (2012).

\bibitem{9_Bettini:2012fu}
A.~Bettini,
\newblock Eur. Phys. J. Plus {\bf 127}, 112 (2012).

\bibitem{10_Kuzminov:2012fv}
V.~V. Kuzminov,
\newblock Eur. Phys. J. Plus {\bf 127}, 113 (2012).

\bibitem{11_Bettini:2012fw}
A.~Bettini,
\newblock Eur. Phys. J. Plus {\bf 127}, 114 (2012).

\bibitem{12_Enqvist:2005cy}
T.~Enqvist {\em et~al.},
\newblock Nucl. Instrum. Meth. {\bf A554}, 286 (2005), hep-ex/0506032.

\bibitem{13_Bandac:2017jjm}
I.~Bandac {\em et~al.},
\newblock J. Phys. Conf. Ser. {\bf 934}, 012019 (2017).

\bibitem{14_Gray:2010nc}
F.~E. Gray {\em et~al.},
\newblock Nucl. Instrum. Meth. {\bf A638}, 63 (2011), 1007.1921.

\bibitem{15_Yu-Cheng:2013iaa}
Y.-C. Wu {\em et~al.},
\newblock Chin. Phys. {\bf C37}, 086001 (2013), 1305.0899.

\bibitem{16_Robinson:2003zj}
M.~Robinson {\em et~al.},
\newblock Nucl. Instrum. Meth. {\bf A511}, 347 (2003), hep-ex/0306014.

\bibitem{17_Blyth:2015nha}
Aberdeen Tunnel Experiment, S.~C. Blyth {\em et~al.},
\newblock Phys. Rev. {\bf D93}, 072005 (2016), 1509.09038,
\newblock [Addendum: Phys. Rev.D94,no.9,099906(2016)].

\bibitem{18_Kalousis:2014wba}
L.~N. Kalousis, E.~Guarnaccia, J.~M. Link, C.~Mariani, and R.~Pelkey,
\newblock JINST {\bf 9}, P08010 (2014), 1406.2641.

\bibitem{19_Esch:2004zj}
E.-I. Esch {\em et~al.},
\newblock Nucl. Instrum. Meth. {\bf A538}, 516 (2005), astro-ph/0408486.

\bibitem{Schmidt:2013gdc}
EDELWEISS, B.~Schmidt {\em et~al.},
\newblock Astropart. Phys. {\bf 44}, 28 (2013), 1302.7112.

\bibitem{20_Ianni:2016fjt}
A.~Ianni,
\newblock J. Phys. Conf. Ser. {\bf 718}, 042030 (2016).

\bibitem{alos}
{ALOS Global Digital Surface Model (AW3D30)},
\newblock {\url{https://www.eorc.jaxa.jp/ALOS/en/aw3d30/}},
\newblock Accessed: 2019-01-09.

\bibitem{21_Kuusiniemi:2018vbz}
P.~Kuusiniemi {\em et~al.},
\newblock Astropart. Phys. {\bf 102}, 67 (2018).

\bibitem{22_Yanin:2011fh}
A.~F. Yanin {\em et~al.},
\newblock (2011), 1101.4489.

\bibitem{Volchenko}
V.~Volchenko {\em et~al.},
\newblock Astrophys. Space Sci. Trans. {\bf 7}, 171 (2011).

\bibitem{geant4}
GEANT4, S.~Agostinelli {\em et~al.},
\newblock Nucl. Instrum. Meth. {\bf A506}, 250 (2003).

\bibitem{24_Lipari}
P.~Lipari and S.~T.,
\newblock Phys. Rev. {\bf D44}, 3543 (1991).

\bibitem{25_Formaggio:2004ge}
J.~A. Formaggio and C.~J. Martoff,
\newblock Ann. Rev. Nucl. Part. Sci. {\bf 54}, 361 (2004).

\bibitem{Mei:2005gm}
D.~Mei and A.~Hime,
\newblock Phys. Rev. {\bf D73}, 053004 (2006), astro-ph/0512125.

\end{thebibliography}

\end{document}